\documentclass[11pt]{article}%
\usepackage{amsfonts}
\usepackage{amsmath}
\usepackage{amssymb}
\usepackage{graphicx}
\usepackage{fullpage}%
\begin{document}

\title{A Brief Tutorial on the Ensemble Kalman Filter\thanks{This document is not
copyrighted and its use is governed by the GNU Free Documentation License,
available at http://www.gnu.org/copyleft/fdl.html. 
%The \LaTeX\ source of this
%document is available at http://www.math.cudenver.edu/\~{}%
%jmandel/papers/enkf\_tutorial . 
The Wikipedia article ``Ensemble Kalman
Filter'' at http://en.wikipedia.org/wiki/Ensemble\_Kalman\_filter as of 06:27,
23 February 2007 (UTC) was created by translating the version of this
 document at that time from
\LaTeX\ to Wiki. This work has been supported by the National Science
Foundation under the grant CNS-0325314, CNS-0719641, and ATM-0835579.}}
\author{Jan Mandel\thanks{Center for Computational Mathematics, University of Colorado
Denver, Denver, CO 80217-3364}}
\date{February 2007, updated January 2009}
\maketitle

\begin{abstract}
The ensemble Kalman filter (EnKF) is a recursive filter suitable for problems
with a large number of variables, such as discretizations of partial
differential equations in geophysical models. The EnKF originated as a version
of the Kalman filter for large problems (essentially, the covariance matrix is
replaced by the sample covariance), and it is now an important data
assimilation component of ensemble forecasting. EnKF is related to the
particle filter (in this context, a particle is the same thing as an ensemble
member) but the EnKF makes the assumption that all probability distributions
involved are Gaussian. This article briefly describes the derivation and
practical implementation of the basic version of EnKF, and reviews several extensions.

\end{abstract}

\section{Introduction}

The Ensemble Kalman Filter (EnKF) is a Monte-Carlo implementation of the
Bayesian update problem:\ Given a probability density function (pdf) of the
state of the modeled system (the \emph{prior}, called often the forecast in
geosciences) and the data likelihood, the Bayes theorem is used to to obtain
pdf after the data likelihood has beed taken into account (the
\emph{posterior}, often called the analysis). This is called a Bayesian
update. The Bayesian update is combined with advancing the model in time,
incorporating new data from time to time. The original Kalman Filter
\cite{Kalman-1960-NAL} assumes that all pdfs are Gaussian (the Gaussian
assumption) and provides algebraic formulas for the change of the mean and
covariance by the Bayesian update, as well as a formula for advancing the
covariance matrix in time provided the system is linear. However, maintaining
the covariance matrix is not feasible computationally for high-dimensional
systems. For this reason, EnKFs were developed
\cite{Evensen-1994-SDA,Houtekamer-1998-DAE}. EnKFs represent the distribution
of the system state using a random sample, called an ensemble, and replace the
covariance matrix by the sample covariance computed from the ensemble. One
advantage of EnKFs is that advancing the pdf in time is achieved by simply
advancing each member of the ensemble. For a survey of EnKF and related data
assimilation techniques, see~\cite{Evensen-2007-DAE}.
%A version of this article is
%also available as the technical report~\cite{Mandel-2007-BTE}.

\section{A derivation of the EnKF}

\subsection{The Kalman Filter}

Let us review first the Kalman filter. Let $\mathbf{x}$ denote the
$n$-dimensional state vector of a model, and assume that it has Gaussian
probability distribution with mean $\mathbf{\mu}$ and covariance $Q$, i.e.,
its pdf is%
\[
p(\mathbf{x})\propto\exp\left(  -\frac{1}{2}(\mathbf{x}-\mathbf{\mu
})^{\mathrm{T}}Q^{-1}(\mathbf{x}-\mathbf{\mu})\right)  .
\]
Here and below, $\propto$ means proportional; a pdf is always scaled so that
its integral over the whole space is one. This probability distribution,
called the \emph{prior}, was evolved in time by running the model and now is
to be updated to account for new data. It is natural to assume that the error
distribution of the data is known; data have to come with an error estimate,
otherwise they are meaningless. Here, the data $\mathbf{d}$ is assumed to have
Gaussian pdf with covariance $R$ and mean $H\mathbf{x}$, where $H$ is the
so-called the observation matrix. The covariance matrix $R$ describes the
estimate of the error of the data; if the random errors in the entries of the
data vector $\mathbf{d}$ are independent, $R$ is diagonal and its diagonal
entries are the squares of the standard deviation (\textquotedblleft error
size\textquotedblright) of the error of the corresponding entries of the data
vector $\mathbf{d}$. The value $H\mathbf{x}$ is what the value of the data
would be for the state $\mathbf{x}$ in the absence of data errors. Then the
probability density $p(\mathbf{d}|\mathbf{x})$ of the the data $\mathbf{d}$
conditional of the system state $\mathbf{x}$, called the data likelihood, is%
\[
p\left(  \mathbf{d}|\mathbf{x}\right)  \propto\exp\left(  -\frac{1}%
{2}(\mathbf{d}-H\mathbf{x})^{\mathrm{T}}R^{-1}(\mathbf{d}-H\mathbf{x})\right)
.
\]

The pdf of the state and the data likelihood are combined to give the new
probability density of the system state $\mathbf{x}$ conditional on the value
of the data $\mathbf{d}$ (the \emph{posterior}) by the Bayes theorem,%
\[
p\left(  \mathbf{x}|\mathbf{d}\right)  \propto p\left(  \mathbf{d}%
|\mathbf{x}\right)  p(\mathbf{x}).
\]
The data $\mathbf{d}$ is fixed once it is received, so denote the posterior
state by $\mathbf{\hat{x}}$ instead of $\mathbf{x}|\mathbf{d}$ and the
posterior pdf by $p\left(  \mathbf{\hat{x}}\right)  $. It can be shown by
algebraic manipulations~\cite{Anderson-1979-OF} that the posterior pdf is also
Gaussian,%
\[
p\left(  \mathbf{\hat{x}}\right)  \propto\exp\left(  -\frac{1}{2}%
(\mathbf{\hat{x}}-\mathbf{\hat{\mu}})^{\mathrm{T}}\hat{Q}^{-1}(\mathbf{\hat
{x}}-\mathbf{\hat{\mu}})\right)  ,
\]
with the posterior mean $\mathbf{\hat{\mu}}$ and covariance $\hat{Q}$ given by
the Kalman update formulas%
\[
\mathbf{\hat{\mu}}=\mathbf{\mu}+K\left(  \mathbf{d}-H\mathbf{\mu}\right)
,\quad\hat{Q}=\left(  I-KH\right)  Q,
\]
where%
\[
K=QH^{\mathrm{T}}\left(  HQH^{\mathrm{T}}+R\right)  ^{-1}%
\]
is the so-called Kalman gain matrix.

\subsection{The Ensemble Kalman Filter}

The EnKF is a Monte Carlo approximation of the Kalman filter, which avoids
evolving the covariance matrix of the pdf of the state vector $\mathbf{x}$.
Instead, the distribution is represented by a collection of realizations,
called an ensemble. So, let
\[
X=\left[  \mathbf{x}_{1},\ldots,\mathbf{x}_{N}\right]  =\left[  \mathbf{x}%
_{i}\right]
\]
be an $n\times N$ matrix whose columns are a sample from the prior
distribution. The matrix $X$ is called the \emph{prior ensemble}. Replicate
the data $\mathbf{d}$ into an $m\times N$ matrix
\[
D=\left[  \mathbf{d}_{1},\ldots,\mathbf{d}_{N}\right]  =\left[  \mathbf{d}%
_{i}\right]
\]
so that each column $\mathbf{d}_{i}$ consists of the data vector $\mathbf{d}$
plus a random vector from the $n$-dimensional normal distribution $N(0,R)$.
Then the columns of
\[
\hat{X}=X+K(D-HX)
\]
form a random sample from the posterior distribution. The EnKF is now obtained
\cite{Johns-2008-TEK} simply by replacing the state covariance $Q$ in Kalman
gain matrix $K=QH^{\mathrm{T}}\left(  HQH^{\mathrm{T}}+R\right)  ^{-1}$ by the
sample covariance $C$ computed from the ensemble members (called the
\emph{ensemble covariance}).

\section{Theoretical analysis}

It is commonly stated that the ensemble is a sample (that is, independent
identically distributed random variables, i.i.d.) and its probability
distribution is represented by the mean and covariance, thus assuming that the
ensemble is normally distributed. Although the resulting analyses, e.g.,
\cite{Burgers-1998-ASE}, played an important role in the development of EnKF,
both statements are false. The ensemble covariance is computed from all
ensemble members together, which introduces dependence, and the EnKF formula
is a nonlinear function of the ensemble, which destroys the normality of the
ensemble distribution. For a mathematical proof of the convergence of the EnKF
in the limit for large ensembles to the Kalman filer, see
\cite{Mandel-2009-CEK}. The core of the analysis is a proof that large
ensembles in the EnKF are, in fact, nearly i.i.d. and nearly normal.

\section{Implementation}

\subsection{Basic formulation}

Here we follow~\cite{Burgers-1998-ASE,Evensen-2003-EKF,Mandel-2006-EIE}.
Suppose the ensemble matrix $X$ and the data matrix $D$ are as above. The
ensemble mean and the covariance are%
\[
E\left(  X\right)  =\frac{1}{N}\sum_{k=1}^{N}\mathbf{x}_{k},\quad
C=\frac{AA^{T}}{N-1},
\]
where%
\[
A=X-E\left(  X\right)  =X-\frac{1}{N}\left(  X\mathbf{e}_{N\times1}\right)
\mathbf{e}_{1\times N},
\]
and $e$ denotes the matrix of all ones of the indicated size.

The posterior ensemble $X^{p}$ is then given by
\[
\hat{X}\approx X^{p}=X+CH^{T}\left(  HCH^{T}+R\right)  ^{-1}(D-HX),
\]
where the perturbed data matrix $D$ is as above. It can be shown that
\emph{the posterior ensemble consists of linear combinations of members of the
prior ensemble}.

Note that since $R$ is a covariance matrix, it is always positive semidefinite
and usually positive definite, so the inverse above exists and the formula can
be implemented by the Choleski decomposition~\cite{Mandel-2006-EIE}.
In~\cite{Burgers-1998-ASE,Evensen-2003-EKF}, $R$ is replaced by the sample
covariance $DD^{T}/\left(  N-1\right)  $ and the inverse is replaced by a
pseudoinverse, computed using the Singular Values Decomposition (SVD).

Since these formulas are matrix operations with dominant Level 3
operations~\cite{Golub-1989-MAC}, they are suitable for efficient
implementation using software packages such as LAPACK (on serial and shared
memory computers)\ and ScaLAPACK (on distributed memory computers)
\cite{Mandel-2006-EIE}. Instead of computing the inverse of a matrix and
multiplying by it, it is much better (several times cheaper and also more
accurate) to compute the Choleski decomposition of the matrix and treat the
multiplication by the inverse as solution of a linear system with many
simultaneous right-hand sides~\cite{Golub-1989-MAC}.

\subsection{Observation matrix-free implementation}

It is usually inconvenient to construct and operate with the matrix $H$
explicitly; instead, a function $h(x)$ of the form
\begin{equation}
h(\mathbf{x})=H\mathbf{x},\label{eq:obs-function}%
\end{equation}
is more natural to compute. The function $h$ is called the \emph{observation
function} or, in the inverse problems context, the \emph{forward operator}.
The value of $h(\mathbf{x})$ is what the value of the data would be for the
state $\mathbf{x}$ assuming the measurement is exact. Then
\cite{Mandel-2006-EIE,Mandel-2007-DAW} the posterior ensemble can be rewritten
as
\[
X^{p}=X+\frac{1}{N-1}A\left(  HA\right)  ^{T}P^{-1}(D-HX)
\]
where%
\[
HA=HX-\frac{1}{N}\left(  \left(  HX\right)  \mathbf{e}_{N\times1}\right)
\mathbf{e}_{1\times N},
\]
and%
\[
P=\frac{1}{N-1}HA\left(  HA\right)  ^{T}+R,
\]
with%
\begin{align*}
\left[  HA\right]  _{i} &  =H\mathbf{x}_{i}-H\frac{1}{N}\sum_{j=1}%
^{N}\mathbf{x}_{j}\\
&  =h\left(  \mathbf{x}_{i}\right)  -\frac{1}{N}\sum_{j=1}^{N}h\left(
\mathbf{x}_{j}\right)  .
\end{align*}
Consequently, the ensemble update can be computed by evaluating the
observation function $h$ on each ensemble member once and the matrix $H$ does
not need to be known explicitly. This formula also holds
\cite{Mandel-2006-EIE} for an observation function $h(\mathbf{x}%
)=H\mathbf{x+f}$ with a fixed offset $\mathbf{f}$, which also does not need to
be known explicitly. The above formula has been commonly used for a nonlinear
observation function $h$, such as the position of a hurricane
vortex~\cite{Chen-2007-AVP}. In that case, the observation function is
essentially approximated by a linear function from its values at the ensemble members.

\subsection{Implementation for a large number of data points}

For a large number $m$ of data points, the multiplication by $P^{-1}$ becomes
a bottleneck. The following alternative
formula~\cite{Mandel-2006-EIE,Mandel-2007-DAW} is advantageous when the number
of data points $m$ is large (such as when assimilating gridded or pixel data)
and the data error covariance matrix $R$ is diagonal (which is the case when
the data errors are uncorrelated), or cheap to decompose (such as banded due
to limited covariance distance). Using the Sherman-\allowbreak
Morrison-\allowbreak Wood\-bury formula~\cite{Hager-1989-UIM}
\[
(R+UV^{T})^{-1}=R^{-1}-R^{-1}U(I+V^{T}R^{-1}U)^{-1}V^{T}R^{-1},
\]
with
\[
U=\frac{1}{N-1}HA,\quad V=HA,
\]
gives%
\begin{align*}
P^{-1} &  =\left(  R+\frac{1}{N-1}HA\left(  HA\right)  ^{T}\right)  ^{-1}\\
&  =R^{-1}\left[  I-\frac{1}{N-1}\left(  HA\right)  \left(  I+\left(
HA\right)  ^{T}R^{-1}\frac{1}{N-1}\left(  HA\right)  \right)  ^{-1}\left(
HA\right)  ^{T}R^{-1}\right]  ,
\end{align*}
which requires only the solution of systems with the matrix $R$ (assumed to be
cheap) and of a system of size $N$ with $m$ right-hand sides. See
\cite{Mandel-2006-EIE} for operation counts.

\section{Further extensions}

The EnKF version described here involves randomization of data. For filters
without randomization of data, see
\cite{Anderson-2001-EAK,Evensen-2004-SSR,Tippett-2003-ESR}.

Since the ensemble covariance is rank-deficient (there are many more state
variables, typically millions, than the ensemble members, typically less than
a hundred), it has large terms for pairs of points that are spatially distant.
Since in reality the values of physical fields at distant locations are not
that much correlated, the covariance matrix is tapered off artificially based
on the distance, which results in better approximation of the covariance for
small ensembles \cite{Furrer-2007-EHP}, such as typically used in practice.
Further development of this idea gives rise to localized EnKF algorithms
\cite{Anderson-2003-LLS,Ott-2003-LEK}.

For problems with coherent features, such as firelines, squall lines, and rain
fronts, there is a~need to adjust the simulation state by distorting the state
in space as well as by an additive correction to the state. The morphing EnKF
\cite{Beezley-2008-MEK,Mandel-2006-PME} employs intermediate states, obtained
by techniques borrowed from image registration and morphing, instead of linear
combinations of states.

EnKFs rely on the Gaussian assumption, though they are of course used in
practice for nonlinear problems, where the Gaussian assumption is not
satisfied. Related filters attempting to relax the Gaussian assumption in EnKF
while preserving its advantages include filters that fit the state pdf with
multiple Gaussian kernels~\cite{Anderson-1999-MCI}, filters that approximate
the state pdf by Gaussian mixtures~\cite{Bengtsson-2003-NFE}, a variant of the
particle filter with computation of particle weights by density estimation
\cite{Mandel-2006-PME,Mandel-2009-EKP}, and a variant of the particle filter
with thick tailed pdfs to alleviate particle filter
degeneracy~\cite{vanLeeuwen-2003-VMF}.

\bibliographystyle{siam}
\bibliography{../../bibliography/dddas-jm}

\def\cprime{$'$} \def\cprime{$'$}
\begin{thebibliography}{10}

\bibitem{Anderson-1979-OF}
{\sc B.~D.~O. Anderson and J.~B. Moore}, {\em Optimal filtering},
  Prentice-Hall, Englewood Cliffs, N.J., 1979.

\bibitem{Anderson-2001-EAK}
{\sc J.~L. Anderson}, {\em An ensemble adjustment {K}alman filter for data
  assimilation}, Monthly Weather Review, 129 (1999), pp.~2884--2903.

\bibitem{Anderson-2003-LLS}
\leavevmode\vrule height 2pt depth -1.6pt width 23pt, {\em A local least
  squares framework for ensemble filtering}, Monthly Weather Review, 131
  (2003), pp.~634--642.

\bibitem{Anderson-1999-MCI}
{\sc J.~L. Anderson and S.~L. Anderson}, {\em A {M}onte {Ca}rlo implementation
  of the nonlinear filtering problem to produce ensemble assimilations and
  forecasts}, Monthly Weather Review, 127 (1999), pp.~2741--2758.

\bibitem{Beezley-2008-MEK}
{\sc J.~D. Beezley and J.~Mandel}, {\em Morphing ensemble {K}alman filters},
  Tellus, 60A (2008), pp.~131--140.

\bibitem{Bengtsson-2003-NFE}
{\sc T.~Bengtsson, C.~Snyder, and D.~Nychka}, {\em Toward a nonlinear ensemble
  filter for high dimensional systems}, Journal of Geophysical Research -
  Atmospheres, 108(D24) (2003), pp.~STS 2--1--10.

\bibitem{Burgers-1998-ASE}
{\sc G.~Burgers, P.~J. van Leeuwen, and G.~Evensen}, {\em Analysis scheme in
  the ensemble {K}alman filter}, Monthly Weather Review, 126 (1998),
  pp.~1719--1724.

\bibitem{Chen-2007-AVP}
{\sc Y.~Chen and C.~Snyder}, {\em Assimilating vortex position with an ensemble
  {K}alman filter}, Monthly Weather Review, 135 (2007), pp.~1828--1845.

\bibitem{Evensen-1994-SDA}
{\sc G.~Evensen}, {\em Sequential data assimilation with nonlinear
  quasi-geostrophic model using {M}onte {C}arlo methods to forecast error
  statistics}, Journal of Geophysical Research, 99 (C5) (1994), pp.~143--162.

\bibitem{Evensen-2003-EKF}
{\sc G.~Evensen}, {\em The ensemble {K}alman filter: {T}heoretical formulation
  and practical implementation}, Ocean Dynamics, 53 (2003), pp.~343--367.

\bibitem{Evensen-2004-SSR}
\leavevmode\vrule height 2pt depth -1.6pt width 23pt, {\em Sampling strategies
  and square root analysis schemes for the {EnKF}}, Ocean Dynamics, 54 (2004),
  pp.~539--560.

\bibitem{Evensen-2007-DAE}
{\sc G.~Evensen}, {\em Data assimilation: {T}he ensemble {K}alman filter},
  Springer, Berlin, 2007.

\bibitem{Furrer-2007-EHP}
{\sc R.~Furrer and T.~Bengtsson}, {\em Estimation of high-dimensional prior and
  posterior covariance matrices in {K}alman filter variants}, J. Multivariate
  Anal., 98 (2007), pp.~227--255.

\bibitem{Golub-1989-MAC}
{\sc G.~H. Golub and C.~F.~V. Loan}, {\em Matrix Computations}, Johns Hopkins
  Univ. Press, 1989.
\newblock Second Edition.

\bibitem{Hager-1989-UIM}
{\sc W.~W. Hager}, {\em Updating the inverse of a matrix}, SIAM Rev., 31
  (1989), pp.~221--239.

\bibitem{Houtekamer-1998-DAE}
{\sc P.~Houtekamer and H.~L. Mitchell}, {\em Data assimilation using an
  ensemble {K}alman filter technique}, Monthly Weather Review, 126 (1998),
  pp.~796--811.

\bibitem{Johns-2008-TEK}
{\sc C.~J. Johns and J.~Mandel}, {\em A two-stage ensemble {K}alman filter for
  smooth data assimilation}, {E}nvironmental and Ecological Statistics, 15
  (2008), pp.~101--110.

\bibitem{Kalman-1960-NAL}
{\sc R.~E. Kalman}, {\em A new approach to linear filtering and prediction
  problems}, Transactions of the ASME -- Journal of Basic Engineering, Series
  D, 82 (1960), pp.~35--45.

\bibitem{Mandel-2006-EIE}
{\sc J.~Mandel}, {\em Efficient implementation of the ensemble {K}alman
  filter}.
\newblock CCM Report 231, University of Colorado Denver, 2006.

\bibitem{Mandel-2006-PME}
{\sc J.~Mandel and J.~D. Beezley}, {\em Predictor-corrector and morphing
  ensemble filters for the assimilation of sparse data into high dimensional
  nonlinear systems}.
\newblock CCM Report 239, University of Colorado at Denver and Health Sciences
  Center. http://www.math.cudenver.edu/ccm/reports/rep239.pdf, November 2006.
\newblock 11th Symposium on Integrated Observing and Assimilation Systems for
  the Atmosphere, Oceans, and Land Surface (IOAS-AOLS), CD-ROM, Paper 4.12,
  87th American Meterological Society Annual Meeting, San Antonio, TX, January
  2007, http://www.ametsoc.org.

\bibitem{Mandel-2009-EKP}
\leavevmode\vrule height 2pt depth -1.6pt width 23pt, {\em An ensemble
  {K}alman-particle predictor-corrector filter for non-gaussian data
  assimilation}.
\newblock ICCS 2009, Lecture Notes in Computer Science, Springer, to appear,
  2009.
\newblock arXiv:0812.2290, 2008.

\bibitem{Mandel-2007-DAW}
{\sc J.~Mandel, J.~D. Beezley, J.~L. Coen, and M.~Kim}, {\em Data assimilation
  for wildland fires: Ensemble {K}alman filters in coupled atmosphere-surface
  models}.
\newblock arXiv:0712.3965, 2007.

\bibitem{Mandel-2009-CEK}
{\sc J.~Mandel, L.~Cobb, and J.~D. Beezley}, {\em On the convergence of the
  ensemble {K}alman filter}.
\newblock arxiv:0901.2951, 2009.

\bibitem{Ott-2003-LEK}
{\sc E.~Ott, B.~R. Hunt, I.~Szunyogh, A.~V. Zimin, E.~J. Kostelich, M.~Corazza,
  E.~Kalnay, D.~Patil, and J.~A. Yorke}, {\em A local ensemble {K}alman filter
  for atmospheric data assimilation}, Tellus, 56A (2004), pp.~415--428.

\bibitem{Tippett-2003-ESR}
{\sc M.~K. Tippett, J.~L. Anderson, C.~H. Bishop, T.~M. Hamill, and J.~S.
  Whitaker}, {\em Ensemble square root filters}, Monthly Weather Review, 131
  (2003), pp.~1485--1490.

\bibitem{vanLeeuwen-2003-VMF}
{\sc P.~van Leeuwen}, {\em A variance-minimizing filter for large-scale
  applications}, Monthly Weather Review, 131 (2003), pp.~2071--2084.

\end{thebibliography}

\end{document}